# Testing Adam-Gibbs relationship in tapped Granular Packings

Xinyu Ai[1], Houfei Yuan[3], Shuyang Zhang[3], Zhikun Zeng[3], Hanyu Li[1], Chengjie Xia[4,*] and Yujie Wang[1,2,3,†]

[1]*Department of Physics, College of Mathematics and Physics, Chengdu University of Technology, Chengdu 610059, China*

[2]*State Key Laboratory of Geohazard Prevention and Geoenvironment Protection, Chengdu University of Technology, Chengdu 610059, China*

[3]*School of Physics and Astronomy, Shanghai Jiao Tong University, Shanghai 200240, China*

[4]*Shanghai Key Laboratory of Magnetic Resonance, School of Physics and Electronic Science, East China Normal University, Shanghai 200241, China*

**Abstract**

Disordered granular packings share many similarities with supercooled liquids, particularly in the rapid increase of structural relaxation time within a narrow range of temperature or packing fraction. However, it is unclear whether the dynamics of granular materials align with those of their corresponding thermal hard sphere liquids, and the particular influence of friction of a granular system remains largely unexplored. Here, we experimentally study the slow relaxation and the steady state of monodisperse granular sphere packings with X-ray tomography. We first quantify the thermodynamic parameters under the Edwards' ensemble, (i.e., effective temperature and configurational entropy), of granular spheres with varying friction, and measure their characteristic relaxation time during compaction processes. We then demonstrate a unified picture of the relaxation process in granular systems in which the Adam-Gibbs (AG) relationship is generally followed. These results clarify the close relationship between granular materials and the ideal frictionless hard sphere model.

**Introduction**

Granular matter is ubiquitous in nature and daily life. Due to dissipative interactions among granular particles and the negligible thermal agitation energy compared with the gravitational potential energy[1], granular materials ordinarily remain in stable static packing states without external energy injection. Consequently, a granular packing is often regarded as being at "zero temperature" and lacking the ergodicity of exploring the phase space as a thermal system, hence making it difficult to study using traditional equilibrium theories. On the other hand, the packing structure of granular spheres has been considered to be close to that of atomic liquids and glasses[2,3], and granular materials are often regarded as macroscopic models of glass-forming systems[4]. Also, the slow relaxation dynamics and dynamic heterogeneity properties of granular materials under external perturbation bear striking resemblance to those of glass-forming materials[5-7]. For example, the compaction processes of granular packings can be qualitatively fitted using similar slow dynamics laws obeyed by glassy systems[8,9], suggesting that the structural relaxation mechanism of glass systems could in principle be applied to macroscopic and frictional granular materials. Yet, these resemblances of granular and glassy systems are largely phenomenological. The glassy dynamics represents one of the most intriguing problems in condensed matter physics, but the origin of the slow dynamics in granular matter remains unresolved until now. Particularly, a quantitative correspondence between granular systems with different friction coefficients and a frictionless hard sphere liquid has not been truly developed.

We think examining the effective thermodynamic nature underlying the relaxation dynamics of granular packings would help establish a convincing mapping between the granular and glass systems. Among the many theories explaining the origin of slow dynamics, the Adam-Gibbs (AG) relation is the first to highlight a possible thermodynamic origin[10]. By introducing the notion of cooperative rearrangement regions (CRRs), which are the smallest regions where rearrangement can take place independently, the AG relation establishes a direct link between configurational entropy and relaxation time, and thus underscores the vital role of thermodynamics in glass transition. Over the past 30 years, a considerable amount of simulations and experiments have validated the AG relation in various supercooled liquids[11-16]. The physical implications of the AG relationship and CRRs have been elusive until the development of the Random First Order Transition (RFOT) theory by Wolynes, Kirkpatrick and Thirumalai in the 1980s[17]. Within RFOT, the change in free energy during the rearrangement of CRRs (i.e., mosaics) consists a surface term representing the increase (cost) in free energy due to the mismatch at the interfaces of different mosaics, and a bulk term reflecting the decrease (gain) in free energy driven by the change in entropy of the CRR as rearrangements occur[18]. The change in

free energy when CRRs grow can be expressed as:

$$\Delta F = \gamma \xi^\theta - TS_c \xi^d, \tag{1}$$

where $\gamma$ is the surface tension, $\theta$ is the surface dimension, $T$ is temperature, $S_c$ is the change in configuration entropy density, and $\xi$ is the linear size of the mosaics. The typical mosaic size

$$\xi \sim \left(\frac{1}{TS_c}\right)^{\frac{1}{d-\theta}}, \tag{2}$$

can be obtained by setting $\Delta F = 0$. Additionally, the relaxation process in a supercooled liquid is considered as a thermal activation[19,20], leading to a relationship between structural relaxation time and the energy barrier, which is assumed to be proportional to $\xi^\psi$,

$$\ln(\tau) \sim \left(\frac{\xi^\psi}{T}\right). \tag{3}$$

Combining (2) and (3), we obtain the relation

$$\ln(\tau) \sim \left(\frac{1}{TS_c}\right)^{\frac{\psi}{d-\theta}}. \tag{4}$$

Within RFOT, Kirkpatrick and Wolynes propose that the exponents $\theta = \psi = d/2$[17], where $d$ is the spatial dimension, and therefore recover the original AG relation.

Given the success of the AG relation in glassy liquid systems, the question of whether it is still valid in disordered granular materials remains intriguing. The distinctive characteristic of granular packings is their static nature, hence kinetic entropy is inherently absent and the calculation of configurational entropy is more straightforward, compared with the challenges of distinguishing configurational entropy from the kinetic one in a supercooled liquid. Additionally, during a typical experimental procedure of preparing a granular packing[21], the system is reasonably supercooled where the AG relation is believed to be applicable. In fact, our recent study has tested the AG relation in granular material systems based on the entropy obtained within the Edwards volume ensemble[22], but the rigorous relationship between the Edwards entropy and the true configurational entropy of a hard-sphere liquid has not been established. This is particularly unclear since the Edwards entropy depends on friction, a unique property of mechanically stable granular packings, which is irrelevant for a supercooled hard sphere liquid. It therefore remains an open fundamental question whether frictional granular systems share the same thermodynamic origin of slow relaxation as their frictionless thermal counterparts.

In this study, we investigate the structure and the corresponding structural relaxation time of tapped granular packings by X-ray tomography. By calculating the configurational entropy based on the Edwards volume ensemble, we test the applicability of AG relation for four types of granular beads with different roughness. We find that the AG relation is clearly observed in our systems over the reasonably long timescale for the structural relaxation time once the influence of friction is removed by a proper rescaling of the Edwards entropy. This suggests that the underlying assumption of the AG relation that configurational entropy plays a pivotal role in the occurrence of slow dynamics is also valid for granular systems.

**Result**

In order to obtain the dynamical and structural variables of granular packings, we prepare packings consisting of four types of monodisperse beads with varying friction coefficients. We tap the packings at different tap intensities using an electric shaker, and record the packing structures after various number of taps by CT scans. Packings are initially prepared to a loose state with 10 taps at an intensity of $\Gamma = 12g$, where $g$ is the gravitational acceleration constant. The packing fraction eventually reaches a constant value after a large number of taps, representing a steady state. Packing structures at the steady state are also recorded. Each packing consists of 6,000 to 8,000 spheres with a diameter of about $D = 6$ mm, and is prepared in a cylindrical container with an inner diameter of $22.5D$. By analyzing the three-dimensional tomographic images, we obtain structural information about the packing. Experimental details can be found in the Methods section.

**Relaxation process**

To study the slow relaxation process, we fit the evolution of the packing fraction over tap times (i.e., the compaction curves) at different tap intensities with the KWW (Kohlrausch-Williams-Watts) form

$$\phi(t) = \phi_\infty - (\phi_\infty - \phi_0)\exp\left[\left(-\frac{t}{\tau}\right)^\beta\right], \qquad (5)$$

from which we obtain the characteristic relaxation time $\tau$ of the system. $\phi_\infty$ and $\phi_0$ represent the steady-state and the initial packing fraction, respectively. For consistency, we fix the stretching exponent $\beta = 0.7$. The fitting results for four types of beads with different roughness are shown in Fig. 1(a-d). It can be observed that the stationary-state packing fraction decreases as tap intensity $\Gamma$ increases (see Fig. 2(a)), and the relaxation time $\tau$ also decreases with increasing $\Gamma$.

**Edwards ensemble**

We further obtain the thermodynamic variables of the packings by analyzing the stationary-state packings, similar to our previous work[23]. According to the statistical mechanical framework proposed by Edwards and coworkers, the compactivity $\chi$ can calculated from an analogous fluctuation theorem according to:

$$\frac{1}{\chi(\phi)} - \frac{1}{\chi^r} = \int_{\phi^r}^{\phi} \frac{d\varphi}{\varphi^2 \, \text{var}(V)}, \qquad (6)$$

where $\text{var}(V) = \frac{\sigma^2(V)}{m}$, $\sigma^2(V)$ is the variance of reduced volume of Voronoi cell $V = \frac{\sum_m V_{voro}}{\sum_m V_p}$ and $m$ is the number of particles within the coarse-grain spherical region. $\phi^r$ is the packing fraction of the random loose packing (RLP), which are 0.568, 0.587, 0.593, and 0.605, respectively for the four types of spheres with different roughness. We use a cubic polynomial to fit the variance: $\text{var}(V) = 0.3810 - 0.5908\phi - 0.5214\phi^2 + 0.8392\phi^3$ (see Fig. 2(b)), and then perform numerical integration to calculate compactivity, whose relationship with packing fraction is shown in Fig. 2(c). The configurational entropy $S_c$ is calculated from another thermodynamic equation:

$$S_c(\phi) - S_{RCP} = \int_{\phi}^{\phi_{RCP}} \frac{d\varphi}{\varphi^2 \chi(\varphi)}, \qquad (7)$$

where $\chi(\varphi)$ is also fitted with a polynomial function in the numerical integration (Fig. 2(d)). Here, the entropy of the random close packing (RCP) is set to 1.1 as the Shannon entropy estimated by Briscoe *et al.*, in ref. [24].

As shown in Fig. 2(c) and 2(d), these thermodynamic equations of the state vary systematically with friction, while, consistently, the compactivity decreases with increasing packing fraction from infinite at RLP to nearly zero around RCP, and the configurational entropy also decreases with $\phi$. The relationship between $S_c$ and packing fraction is similar to that of a hard sphere glass. The larger $S_c$ for rougher particles is apparent since more configurations can maintain mechanical stability at a given packing fraction if the particles are more frictional.

**Dynamics-structure relationships**

We now examine the relationships between relaxation time $\tau$ and thermodynamic quantities: compactivity $\chi$ and configurational entropy $S_c$. As shown in Fig. 3(a), $\tau$ increases rapidly as $\chi$ decreases, which can be well described by the Vogel-Fulcher-Tamman (VFT) form for glassy systems. Notably, the $\tau - \chi$ relationships for different frictional materials collapsed on the same curve, in contrast to the friction-dependent behaviors of $\tau - \phi$. This suggests a universal dependence of the structural relaxation time on compactivity or tap intensity instead of packing fraction since it is experimentally observed that various systems possess identical compactivity under the same $\Gamma$ [23]. This is likely owing to the fact that during the tap process, the system relaxes mainly during the heated phase corresponding to liquid configurations when particles are colliding without frictional contacts, and the effects of friction only come into play when the system is quenched rapidly to the jammed state when the system can no longer relax[25].

Interestingly, in contrast with the $\tau - \chi$ relationship, the relationships between relaxation time and Edwards entropy are system-dependent, as shown in Fig. 3(c). It is important to note that the entropy we obtain from the Edwards volume ensemble is not equivalent to the configurational entropy of the hard-sphere liquid. This discrepancy arises because the presence of friction significantly enhances the number of mechanically stable states within each free energy basin, while these configurations do not possess a one-to-one correspondence to the liquid configurations when structural relaxation occurs. Therefore, we need to use the configurational entropy associated with the liquid state to test the AG relation. Consistent with our previous study[25], we can linearly rescale the configurational entropy and packing fraction of different systems by defining $\tilde{S} = \frac{S_c - S_{RCP}}{S_{RLP} - S_{RCP}}$ and $\tilde{\phi} = \frac{\phi - \phi_{RLP}}{\phi_{RCP} - \phi_{RLP}}$, where $S_{RLP}$ and $\phi_{RLP}$ are the entropy and packing fraction of RLP. After this rescaling, the $\tilde{S} - \tilde{\phi}$ relationships can be collapsed across different systems (Fig. 3(b)). Since we believe RLP and RCP correspond to the onset temperature $T_{onset}$ and dynamical glass transition temperature $T_d$ of a glass system, we adopt their absolute values from numerical simulations by setting $\tilde{S}_{rlp} = 1.5$ and $\tilde{S}_{rcp} = 0.5$, respectively[26]. The normalized entropy $\tilde{S}$ bears great resemblance in its behavior to that of a frictionless hard sphere liquid between $T_{onset}$ and $T_d$. Based on this, we plot the $\tau - \tilde{S}$ relationships in Fig. 3(d). Surprisingly, again all the curves collapsed onto a master curve. This in turn suggests that the structural relaxation time is controlled only by the corresponding liquid configurational entropy instead of the frictional ones. The master curve closely matches the predictions by the AG theory, appearing as a straight line on the semi-logarithmic plot, confirming that the original AG relationship is recovered.

At last, we discuss whether and how the physical picture of the RFOT theory matches with the structural relaxation of our granular systems. According to Eq. (4), there exist many potentially valid combinations of values of $\psi$ and $\theta$ that can give the correct AG relation. The RFOT theory predicts that $\theta = \psi = d/2$ [17]. Yet, recent experiments[22,27] indicate that $\theta$ is slightly greater than 2, leading to a $\psi$ value lower than that predicted by RFOT. Since $\psi$ is directly related to the potential barrier between the start and end configurations, this relatively small $\psi$ indicates that the energy barrier does not grow as rapidly as the size of the CRR if activation is the only mechanism for structural relaxation. Instead, some other dynamic relaxation mechanisms may also be working in this regime. Additionally, it should be noted that the macroscopic granular packing corresponds to a hard-sphere liquid quenched between $T_{onset}$ and $T_d$. In this temperature range where the system is not deeply supercooled, the AG relationship is not supposed to hold rigorously, as the activation-dominated regime should typically lie below the dynamical transition temperature $T_d$. However, some studies have pointed out that activation processes are already present near the onset temperature $T_{onset}$[20,28]. Furthermore, it has been observed before that close to $T_d$, the structural relaxation displays chain-like or fractal structures in colloids[29]. This could be related to complex relaxation mechanisms, such as dynamic facilitation, which is supposed to dominate in this regime as relaxation events in a certain location can induce further relaxation in neighboring sites[30]. As a result, the activation of the CRR may not be the dominant structural relaxation mechanism, and it should not be taken for granted that the two free exponents obtained in an experimental system coincide with the prediction of RFOT, when the effective temperature is relatively high such as our tapped granular systems. Surprisingly, in spite of the possibly mixed relaxation dynamics, the AG relation still holds.

**Conclusion**

In summary, we systematically investigate granular packing systems under tap, and use the Edwards volume ensemble framework to calculate relevant statistical mechanical quantities. Upon rescaling the configurational entropy, we observe a general recovery of the AG relationship. This implies that the relaxation time is predominantly governed by the configurational entropy of the ideal frictionless hard-sphere liquid. On the other hand, while granular materials are not deeply supercooled, the complex dynamical mechanisms result in discrepancies between the exponents and RFOT predictions. Furthermore, within this temperature range, the energy barrier is sufficiently large compared to thermal fluctuation. Therefore, we consider the effective temperature to be more appropriate for describing the system's state. In granular sys-tems, the thermodynamic quantities calculated using the Edwards ensemble are inherently configurational-dependent, which accounts for the validity of the AG relationship. Nevertheless, universal relaxation processes in granular systems and their great analogy with hard-sphere glassy systems imply that granular materials and hard-sphere systems share the same thermodynamic and dynamical properties[31-33]. We can see that granular materials exhibit highly similar properties to those of glassy systems in both dynamics and statistical mechanics. The significance of granular materials as macroscopic models of glassy systems is profound, which can be used to investigate the complex relaxation process in real space close to the dynamical transition temperature.

Corresponding authors

*cjxia@phy.ecnu.edu.cn

†yujiewang@sjtu.edu.cn

*Acknowledgments:* The work is supported by the National Natural Science Foundation of China (No. 12274292)

*Author contributions*: Y.W. designed the research. X.A., H.Y., S.Z., Z.Z., H.L., C.X., and Y.W. performed the experiment. X.A., H.Y., S.Z., Z.Z., C.X and Y.W. analyzed the data and wrote the paper.

*Competing Interests statement*: The authors declare no competing interests.

**Experiment**

We utilize X-ray tomography to reconstruct the three-dimensional structures of granular packings monodisperse spherical particles, the imaging resolution is 0.2 millimeters (i.e., $1/30D$) per pixel. Random packing structures are prepared systematically via mechanical tap. Each tap is driven by a pulse wave generated by an arbitrary waveform generator (Tek AFG3100). Each tap cycle consists of a 1/30 s pulse followed by a 1.5 s interval allowing the system to settle.

We employ four types of beads: 3D-printed particles fabricated with a 3D printer with a printing accuracy of 0.032 mm (denoted as 3DP), two kinds of acrylonitrile-butadiene-styrene copolymer particles (denoted as ABS1 and ABS2), and BUMP particles with 150 semi-spheres of diameter $0.1D$ decorated on the sphere surface. The first three types of beads have a diameter of 6mm, and the BUMP has an outer diameter of 5.8 mm. In addition, the effective friction coefficients decrease in the order of BUMP, 3DP, ABS2, and ABS1. To avoid crystallization, we decorate the internal walls of the container with 5 mm and 8 mm diameter ABS semi-spheres at random positions. Depending on tap intensity $\Gamma$, the number of taps needed to reach the stationary states ranges from 10 to 6000. Through image analysis[34,35], we are able to obtain the position and radius of each particle, and calculate all structural parameters of packings. By performing Voronoi tessellation on the packing structure, we calculate the global packing fraction, which is defined as $\phi = \sum V_p / \sum V_{voro}$, where $V_{voro}$ is the volume of Voronoi cells and $V_p$ is the volume of the particle. We set $V_p$ to unity for convenience.

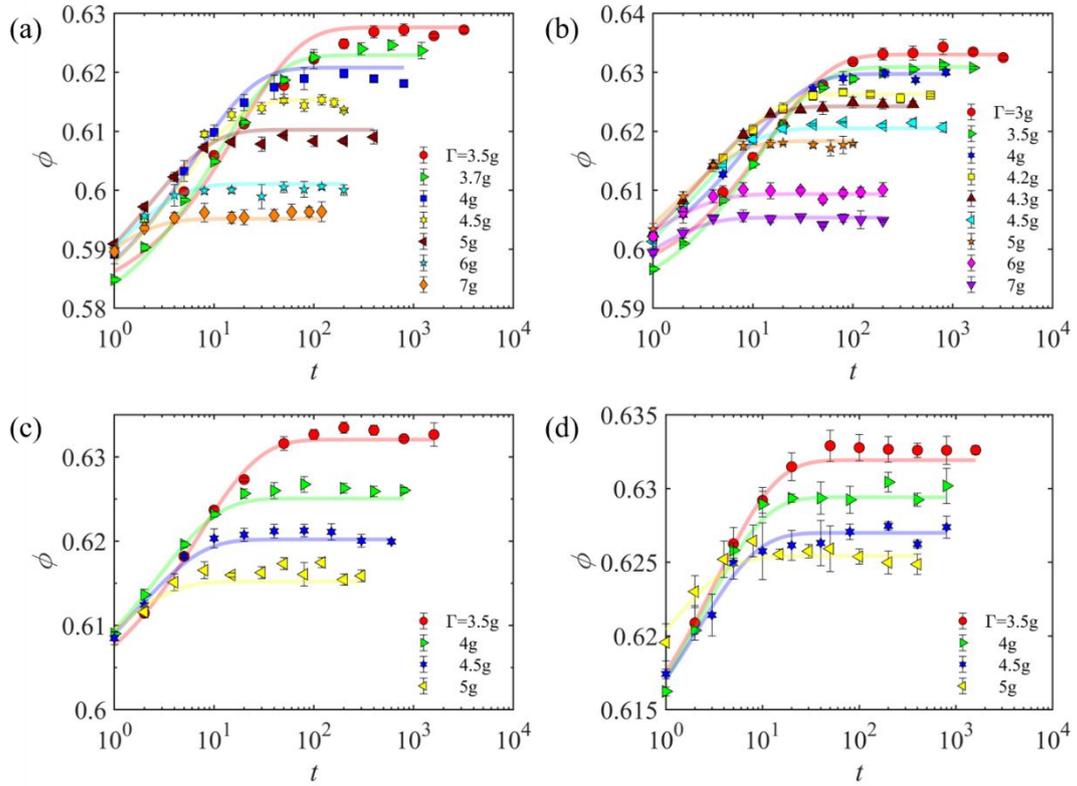

**Fig. 1| Compaction curves of tapped granular packing.** Packing fraction $\phi$ as a function of the number of taps $t$ for **(a)** BUMP, **(b)** 3DP, **(c)** ABS2 and **(d)** ABS1 with different tap intensities $\Gamma$ (different symbols). Solid curves are fittings with Eq. (5). The error bars mark standard deviations of data from three independently repeated experiments. The compaction curves at high tap intensities are not included because the relaxation process is too rapid to accurately measure the relaxation time.

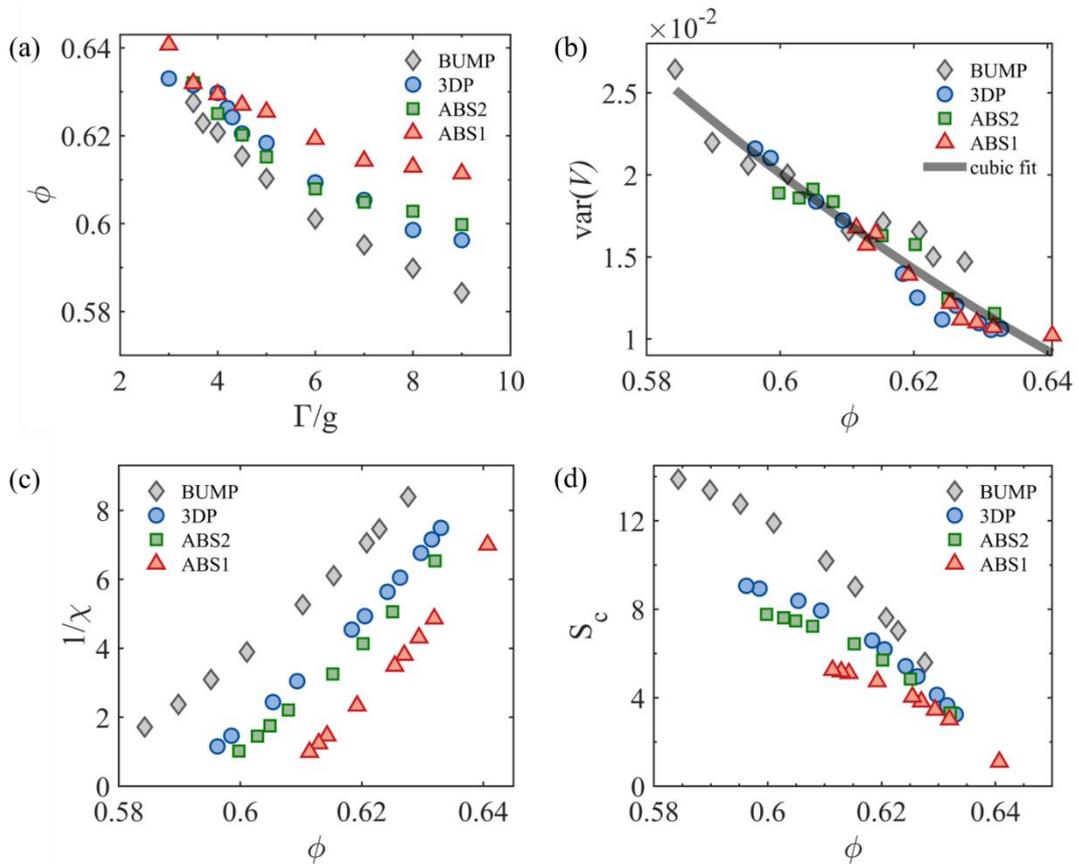

**FIG. 2 | Thermodynamics of various frictional systems. a** Packing fraction $\phi$ as a function of $\Gamma$ for BUMP, 3DP, ABS2, and ABS1 systems. **b** Coarse-grained Voronoi volume variance $\text{var}(V)$ as a function of $\phi$. The solid curve is a cubic polynomial fit (see main text). **c** Inversed compactivity $1/\chi$ as a function of $\phi$. **d** Configurational entropy $S_c$ as a function of $\phi$.

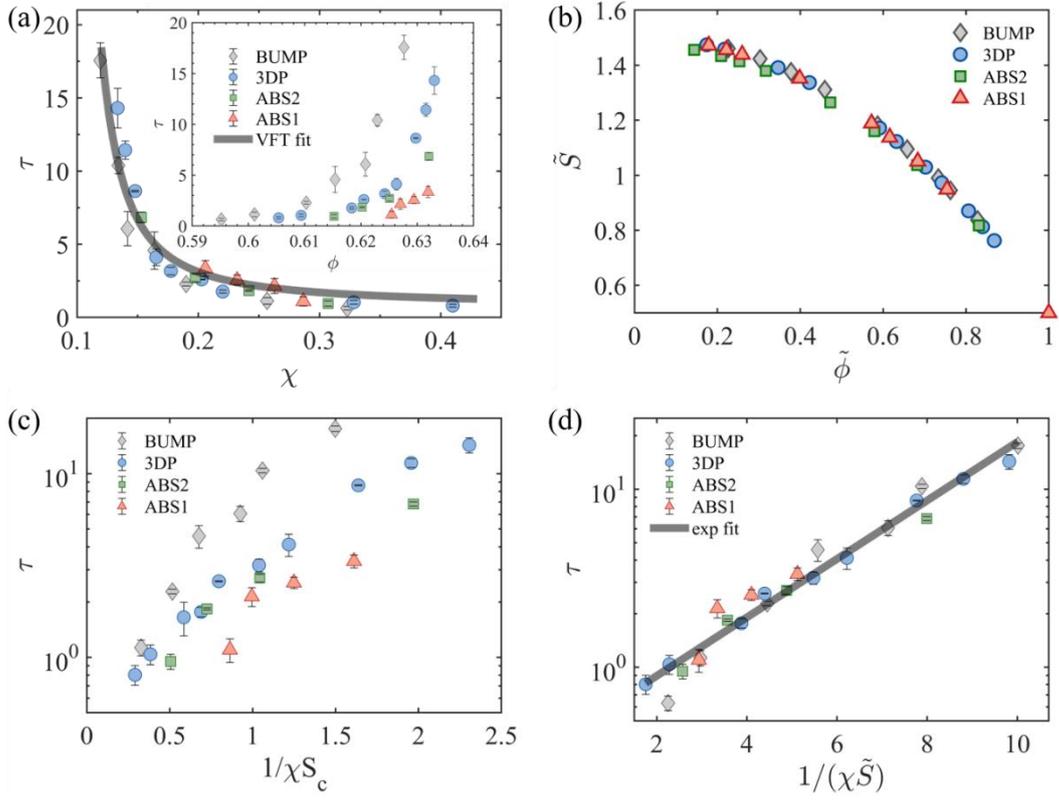

**FIG. 3 | Relationship of dynamics and thermodynamic variables. a** Relaxation time $\tau$ versus $\chi$ for the four systems. The solid curve is a fit of the VFT form $\tau = \tau_0 \exp\left(\dfrac{D\chi_0}{\chi - \chi_0}\right)$. Inset: $\tau$ as a function of $\phi$. **b** Rescaled configurational entropy $\tilde{S}$ versus rescaled packing fraction $\tilde{\phi}$. **c** Relaxation time $\tau$ versus $1/(\chi S_c)$ for five systems. **d** Relaxation time $\tau$ versus $1/(\chi \tilde{S})$. The solid curve is a fit of $\tau \sim \exp[A/(\chi \tilde{S})]$, where $A$ is a constant.